\documentclass[manuscript]{aastex} 
\begin{document}

\shorttitle{Mid-IR Light Curve of SN\,2011dh}
\shortauthors{Helou et al.}

\title{The Mid-Infrared Light Curve of Nearby Core-Collapse Supernova SN\,2011dh (PTF\,11eon)}

\author{George Helou\altaffilmark{1},  Mansi M. Kasliwal\altaffilmark{2}, Eran O. Ofek\altaffilmark{3}, Iair Arcavi\altaffilmark{3}, Jason Surace\altaffilmark{1}, Avishay Gal-Yam\altaffilmark{3}} %Shri Kulkarni\altaffilmark{4}}
\altaffiltext{1}{Infrared Processing and Analysis Center, California Institute of Technology, Pasadena, California 91125}
\altaffiltext{2}{Observatories of the Carnegie Institution for Science, 813 Santa Barbara St, Pasadena, CA, 91101, USA}
\altaffiltext{3}{Benoziyo Center for Astrophysics, Faculty of Physics, The Weizmann Institute of Science, Rehovot 76100, Israel}
%\altaffiltext{4}{Cahill Center for Astrophysics, California Institute of Technology, Pasadena, CA, 91125, USA}

\begin{abstract}
We present Spitzer observations at 3.6 and 4.5$\mu$m of the supernova SN\,2011dh (PTF\,11eon) in M51 from
18\,days to 625\,days after explosion. The mid-infrared emission peaks at 24\,days after explosion 
at a few $\times$10$^7$\,L$_{\odot}$, and decays more slowly than the visible-light bolometric luminosity.  The infrared color temperature cools 
for the first 90\,days and then is constant. Simple numerical models of a thermal echo can qualitatively reproduce the early behavior. 
%Integrating the mid-infrared light curve for the first 45 days places a lower limit on the total energy in the shock breakout flash 
%of 2$\times$10$^{47}$\, ergs.  This corresponds to a shock breakout luminosity greater than 10$^{10}$\,L$_{\odot}$ if the flash lasted
%for less than an hour.  
At late times,  the mid-IR light curve cannot be explained by a simple thermal echo model, suggesting additional dust heating or line emission mechanisms.  We also propose that thermal echoes can serve as effective probes to uncover 
supernovae in heavily obscured environments, and speculate that under the right conditions, integrating the early epoch of the mid-infrared light curve may constrain the total energy in the shock breakout flash.
%%% We fit an illustrative thermal echo model to the light curve to derive the circumstellar density profile and dust properties. 
\end{abstract}


\section{Introduction}

Thermal echoes from supernovae (SN), emanating from preexisting dust 
%Z  in an extended circumstellar envelope 
responding to heating by the shock breakout flash (SBF) of the SN and later by the fireball, are powerful tools to study the dust distribution around the SN, and sometimes the heating sources themselves \citep{w80, dab+10}.  
Light echoes at $<$100\,days would illuminate material from stellar outflows thus constraining progenitor properties. This is particularly valuable for SN types such as IIb whose origins are still under debate. 
%2% (cf. review by \cite{g+11}).
Few cases of mid-infrared (mid-IR) emission from core-collapse SNe  have been studied with well-sampled light-curves and IR spectra, and interpretation has not been straightforward. 
%%% Thermal echoes should be a common occurrence, but separating them from the background emission of the galaxy and following them in time remain challenging.  When measured, the color temperature (or spectra), time scales and relative importance of IR emission can reveal details such as density, spatial distribution and possibly composition  of the pre-existing dust .

\cite{g+11} discuss implications of IR echoes for dust production in SNe.
\cite{krb+05} study a spatially resolved  IR echo associated with late flares in the central object of the SN\,IIb remnant Cas A.
SN1993J (IIb) showed an increasing mid-IR excess beyond 100 days after  explosion \citep{mna+02},  whereas SN1987A (II-pec)
took a longer time to develop that excess \citep{msv+89,msa+93}.
Both SNe 
%Z  SN1993J and SN1987A 
exhibited increasing dust luminosity with time, an indication of the heating reaching more dust mass at later times.
In the case of SN1987A, the visible light echo supports this interpretation \citep{dab+10}. 
Most recently, \cite{f+13} presented mid-IR observations of a sample of SN\,IIn with nearly constant IR excess up to several years post-explosion.

SN\,2011dh, a Type IIb SN in M51, offered a unique opportunity to study its IR properties given the relative proximity of M51 at 8 Mpc, and the location of the SN in the less crowded part of the disk about 6~kpc from the nucleus (\S 2).  
We started monitoring SN\,2011dh with Spitzer shortly after explosion, and we report here on data from almost two years post explosion (\S 3).  We describe the spectral energy distribution (\S 4), discuss  a physical scenario for the emission (\S 5), analyze the data in the framework of that scenario and present a simple illustrative model  (\S 6). Finally, we discuss the implications for the SBF and the utility as a probe to discover SNe  (\S 7).

\section{Discovery and Progenitor constraints}

On May 31.9 2011 UT, the supernova SN\,2011dh (PTF\,11eon) in M51 was discovered (A. Riou) and 
spectroscopically classified as a Type IIb \citep{agy+11}.
Identification of a progenitor star in pre-explosion imaging by the Hubble Space Telescope suggested a yellow supergiant (YSG) progenitor \citep{mfe+11}, although a binary system including a Wolf-Rayet star could not be ruled out  \citep{vlc+11}.  Relatively low temperatures in the earliest spectra, suggestive of rapid shock breakout cooling, pointed to a progenitor radius more compact than a supergiant (\citealt{agy+11} but see \citealt{bbn+12}). Radio and X-ray observations also suggested a compact progenitor \citep{smz+12}.   Later work, taking into account the assumptions in microphysics, demonstrated that the progenitor radius is not well constrained and might be in between the canonical compact and extended classes \citep{hsf+12}. Most recently, \cite{v+13}  reported the YSG to be missing in deep {\it HST} images at 
$\approx$600\,days, strongly supporting its identification as the progenitor, which we adopt  in this paper.
\cite{sac13} argue this is consistent with a binary model.

\section{Spitzer Observations} 

\begin{figure}[!ht] 
\centering
\includegraphics[width=0.7\textwidth]{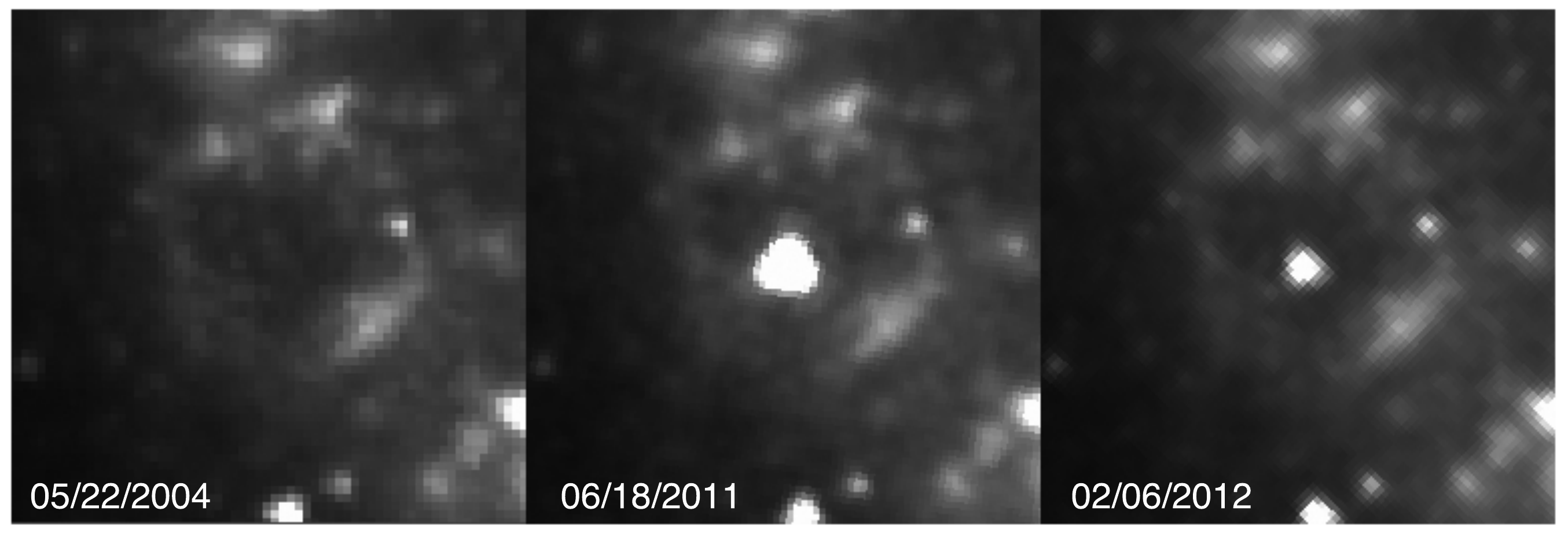}
\caption{Spitzer images of SN\,2011dh  at 3.6$\mu$m, one from the first follow-up epoch (middle frame),
and one about eight months later (right-hand frame). The left-hand frame is a pre-explosion archival reference image.}
\label{fig:image}
\end{figure}

We observed SN\,2011dh  17 times{\footnote{not counting one observation which was ruined by radiation hits}} with Spitzer/IRAC \citep{f+04}, starting  18\,days after explosion and extending to 625\,days after explosion.  Figure~\ref{fig:image} shows three Spitzer images of the supernova:
pre-explosion,  early epoch and  late epoch post-explosion.

Data were processed using the standard IRAC pipeline{\footnote{IRAC Instrument and Instrument Support Teams, IRAC Instrument Handbook version 2.0.3 February 2013}}. 
 Calibration and aperture corrections were as per the zero points listed in the Spitzer/IRAC handbook.  To remove underlying host galaxy 
light, image subtraction was performed using HOTPANTS{\footnote{http://www.astro.washington.edu/users/becker/hotpants.html}} (A. Becker).  Aperture photometry was then performed using a radius of 4 pixels for the supernova and an annulus of 8 to 16 pixel radii for the background.
%%%  Note that the image subtraction results were significantly different from using an annulus to measure background only for data beyond 200\,days.  

%All except the last two epochs have detections with very high signal-to-noise ratio 
%($\gg$30) and photometric uncertainties are dominated by systematics estimated at 5\%. 
%The last two epochs have significant  uncertainty from the removal of underlying host galaxy light. REWRITE FOR SUBTRACTED IMAGE MEASUREMENTS.
%Photometry at days 273 and 383 is corrupted because of radiation hits and other artifacts.

\begin{figure}[!ht] 
\centering
\includegraphics[width=0.75\textwidth]{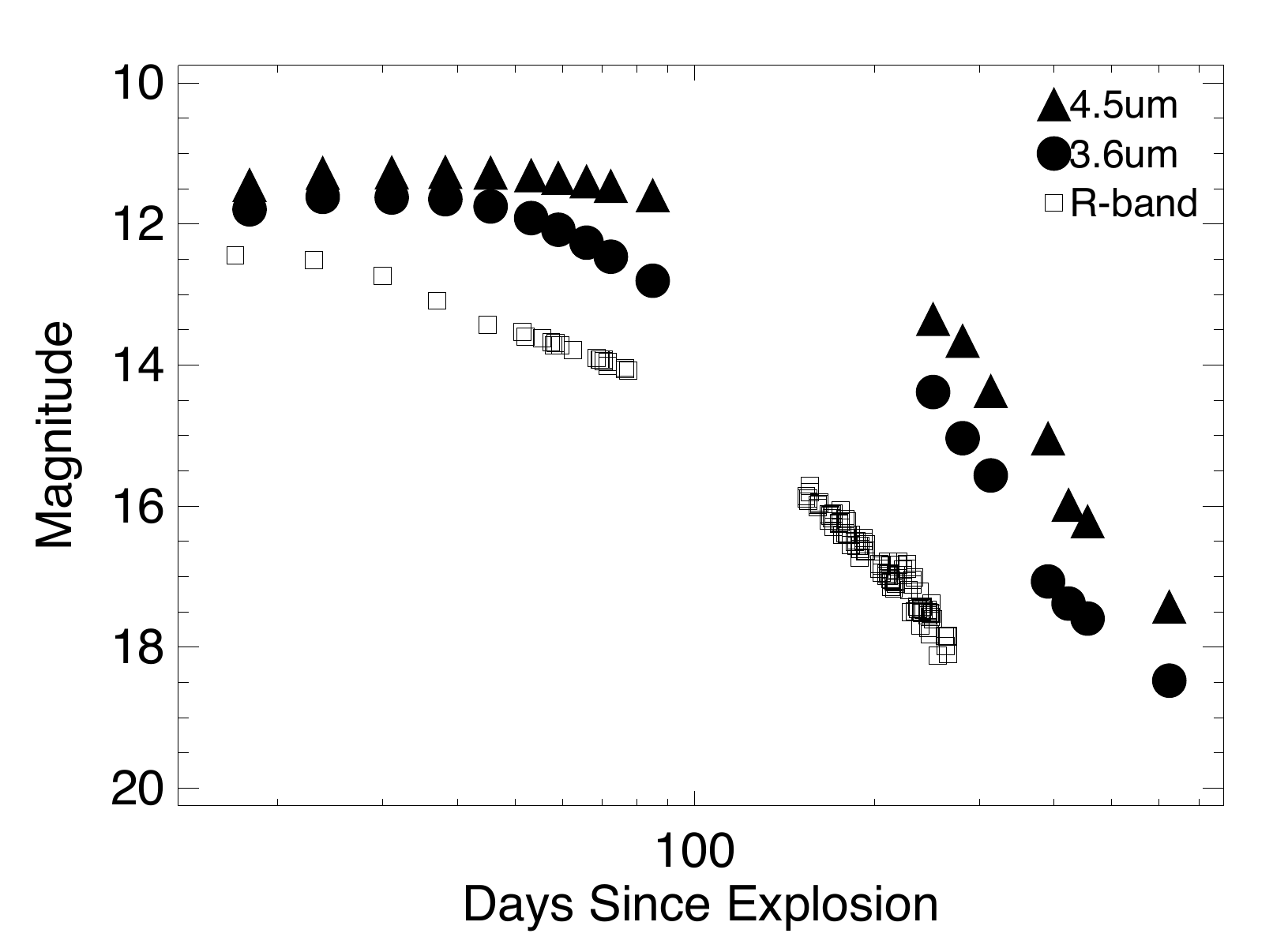}
\caption{SN\,2011dh  light curves at 3.6 and 4.5$\mu$m, shown as filled circles and triangles respectively. 
The $R$-band data are shown as empty squares. Nowhere does the $R$-band decay slope agree exactly with the mid-IR slopes, though all curves show the slow-to-fast decay transition.
}
%The dashed vertical line indicates likely explosion time.
%The V light curve is shown as the open squares. }
\label{fig:lczoom}
\end{figure}

The Spitzer photometry is presented in Figure~\ref{fig:lczoom} and compared to the $R$-band light curve from the Palomar Transient Factory (Arcavi et al. in prep). % and  [TO BE REMOVED]  Figure~\ref{fig:lc}. % and Table~\ref{tab:phot}. 
The light curve peaked 24 to 30\,days after explosion
at 3.6$\mu$m,  and at 38\,days at 4.5$\mu$m, and declined faster at 3.6$\mu$m 
than at 4.5$\mu$m.  The data gap between 90 and 250 days is due to  visibility gaps.  The light curve drops much faster after the gap at all wavelengths, with a possible slow-down beyond 400\,days.
Data interpolation into the gap for 4.5$\mu$m points to transition from plateau to decay  at $150\pm25$\,days.

%%% \begin{figure*}[!ht] 
%%% \centering
%%% \includegraphics[width=\textwidth]{11eon_fulllc.pdf}
%%% \caption{Panchromatic light curve of PTF\,11eon including UV, optical, near-IR and mid-IR photometry.
%%% Early optical and near-infrared data is from \citealt{agy+11} and late-time optical data is from XX in prep.}
%%% \label{fig:lc}
%%% \end{figure*}

\section{Spectral Energy Distribution}

\begin{figure}[!ht] 
\centering
\includegraphics[width=\textwidth]{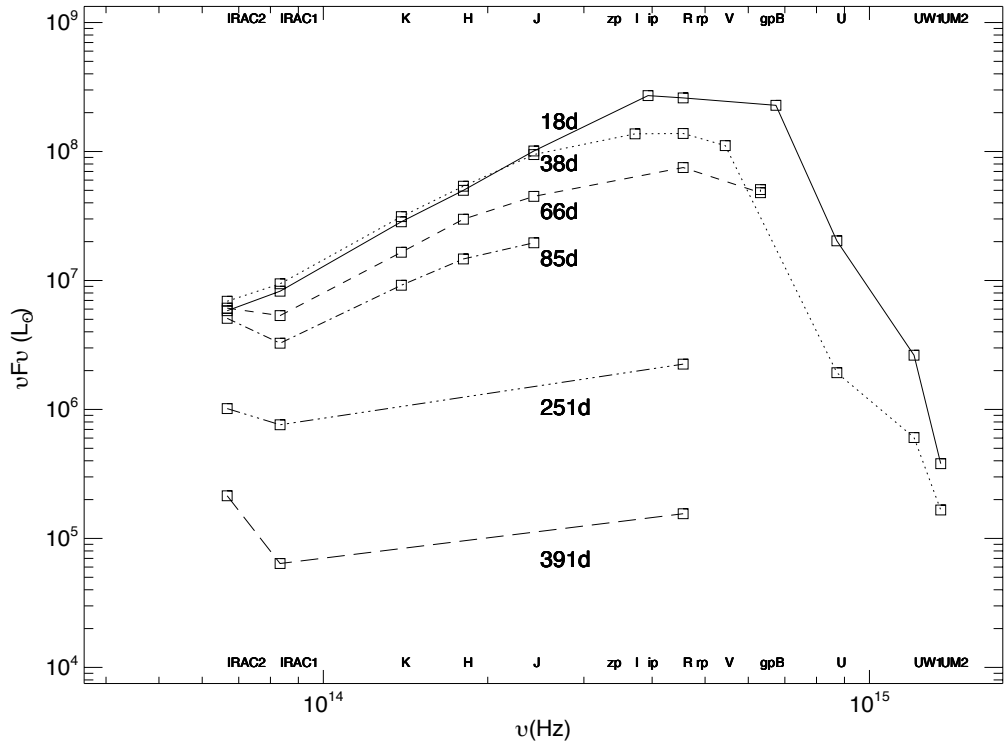}
\caption{SN\,2011dh spectral energy distributions from the ultra-violet to the mid-infrared at six different epochs.
Note that the contribution of the mid-infrared to the bolometric light curve grows from a few \% to 50\%.}
\label{fig:sed}
\end{figure}

In Figure~\ref{fig:sed}, we show the spectral energy distribution (SED) of SN\,2011dh  ranging from the UV
using {\it Swift} satellite data to the IR using Spitzer data. The SED is shown at six epochs between
18\,days and 391\,days after explosion.

SN\,2011dh exhibits substantial  IR emission whose importance    increases with time.
The [R-4.5$\mu$m]  color grows from $1$\,mag to $4.8$\,mag between 18 and 391\,days.
Visible and UV luminosities, and thus bolometric luminosity, decay with a 15-day half-life \citep{agy+11}, whereas the infrared decays more slowly up to 50\,days after explosion.  
The NIR data \citep{e+13} prove critical to deciphering the SED. They show a hot component peaking around $1\mu$m, with the 4.5$\mu$m emission emerging as a separate component after two months. At 66\,days, a simple extrapolation of H and K data on Figure~\ref{fig:sed} accounts for most of the 3.6$\mu$m flux, but for only half of the 4.5$\mu$m flux.  
This is consistent with \cite{e+13} accounting for the NIR as part of a fit to visible data at 5000-7000\,K, and associating the 4.5$\mu$m\, with an additional 400\,K component.  The lack of NIR data complicates the decomposition after 100\,days, but it remains safe to treat at least 4.5$\mu$m as a separate component associated with the IR echo.

Between 50 and 100\,days, the mid-IR flux densities decay with a power-law dependence on time with an index of $\approx\,-0.7$ for 4.5$\mu$m and a faster $\approx\,-1.8$  for 3.6$\mu$m, so that the mid-IR emission grows redder  (Figure~\ref{fig:lczoom}). Beyond 250\,days, the light curves decay essentially in parallel with a steeper  index of $\approx\,-5$.  Nowhere does the $R$-band decay slope agree exactly with the mid-IR slopes, though all curves transition from slow to fast decay.
%%%
The mid-IR color temperature, estimated form a blackbody curve that reproduces the $f_\nu(3.6)/f_\nu(4.5)$ ratio, starts out above 1600\,K then cools slowly, reaching 600\,K about 85 days after explosion. 
Beyond 200\,days, it ranges between 500 and 700\,K, 
%%% with fluctuations consistent with being  due to measurement errors.
but it is unclear how much the hot component  still contributes at 3.6$\mu$m, leaving room for a cooler IR echo.
We adopt conservatively 500\,K as the dust temperature at times $>200$\,days.

% NOTE: The values of -0.7 and -1.8 in the paragraph above are rounded from -0.68 and -1.84 measured by hand

%%% Due to the lack of reference image for the Spitzer images (Mansi - do we have a pre-explosion ref image?), we also removed the last data point.

%%% \section{Physical Picture}

\section{Analysis}

We adopt the physical picture  \citep{w80, d83, gm86, f+10} where the IR emission is due to dust heated first by the SN SBF (a brief, high intensity UV flash) and subsequently by the fireball. We assume the dust  is spherically symmetric about the SN.  The dust contributing to the instantaneous IR light observed at Earth is distributed along the surface of equal time of light travel from SN to dust to observer, outside a dust sublimation radius $r_{sub}$.
%%% SN\,2011dh appears to reach this peak  at 30 to 40\,days after explosion.  
%%% As the flash intensity continues to weaken, the apparent temperature of the mid-infrared emission  cools off, as observed beyond the first month.  
The longer-duration  fireball of shocked gas will heat the dust left or reformed behind the SBF, as well as new dust formed in the ejecta. 
A blackbody approximation is quite unlikely to be appropriate for the observed emission, as the equal light travel time surface spans a  range of distances from the SN. For example,  \citep{f+10}  discuss at least two components in SN2005ip, at about 470 and 900K.  While the cooler component is consistent with 4.5$\mu$m data in SN2011dh, our NIR component is at T$>$2000\,K range, unlikely to be due to dust emission.

The MIR transition to late epochs seems complicated.  
%Z  Trying to connect simply the early and late photometry, one finds that the 4.5$\mu$m light curve 
At 4.5$\mu$m extrapolating early and late photometry into the data gap suggests a light curve  turning over about 150\,days after explosion.  At 3.6$\mu$m extrapolations into the data gap from early and late data do not intersect, implying that the decay must have slowed down for some time  before the steeper decline.  This may be due to the hot NIR component.

\cite{d83} associates the IR echo plateau duration with the inner radius of the dust cloud at $r_{sub}$, suggesting a value of 150\,days or $\approx$0.1\,pc for SN\,2011dh. While the association is not a precise relation in more realistic models, it remains a valid indicative relation.  Reading off Figure 8 of \cite{f+10}, $r_{sub}\approx$0.1\,pc requires an uncomfortably high peak luminosity of $10^{12}L_{\odot}$.  Alternatively, the inner radius may be set by a sharp decline of mass outflow about 150\,days before explosion. Similarly, the \cite{d83} model requires a very extended dust envelope to generate the late epoch emission.   
The latest detections in our data at  $\approx\,625$\,days probe dust  at distances up to  0.26\,pc behind  the SN, and twice as far in the lateral directions.  
Dust condensing in a mass outflow from a YSG  would be the first to echo the SBF in IR. For outflow velocities of 100 $\rm km/s$, the stellar material would have required $<3000$\,years to reach 0.26\,pc,  consistent with typical estimates of outflow durations.

\subsection{Energy and Mass Estimates}

%%% \subsection{Thermal Balance Case}

\begin{figure}[!ht] 
\centering
\includegraphics[width=\textwidth]{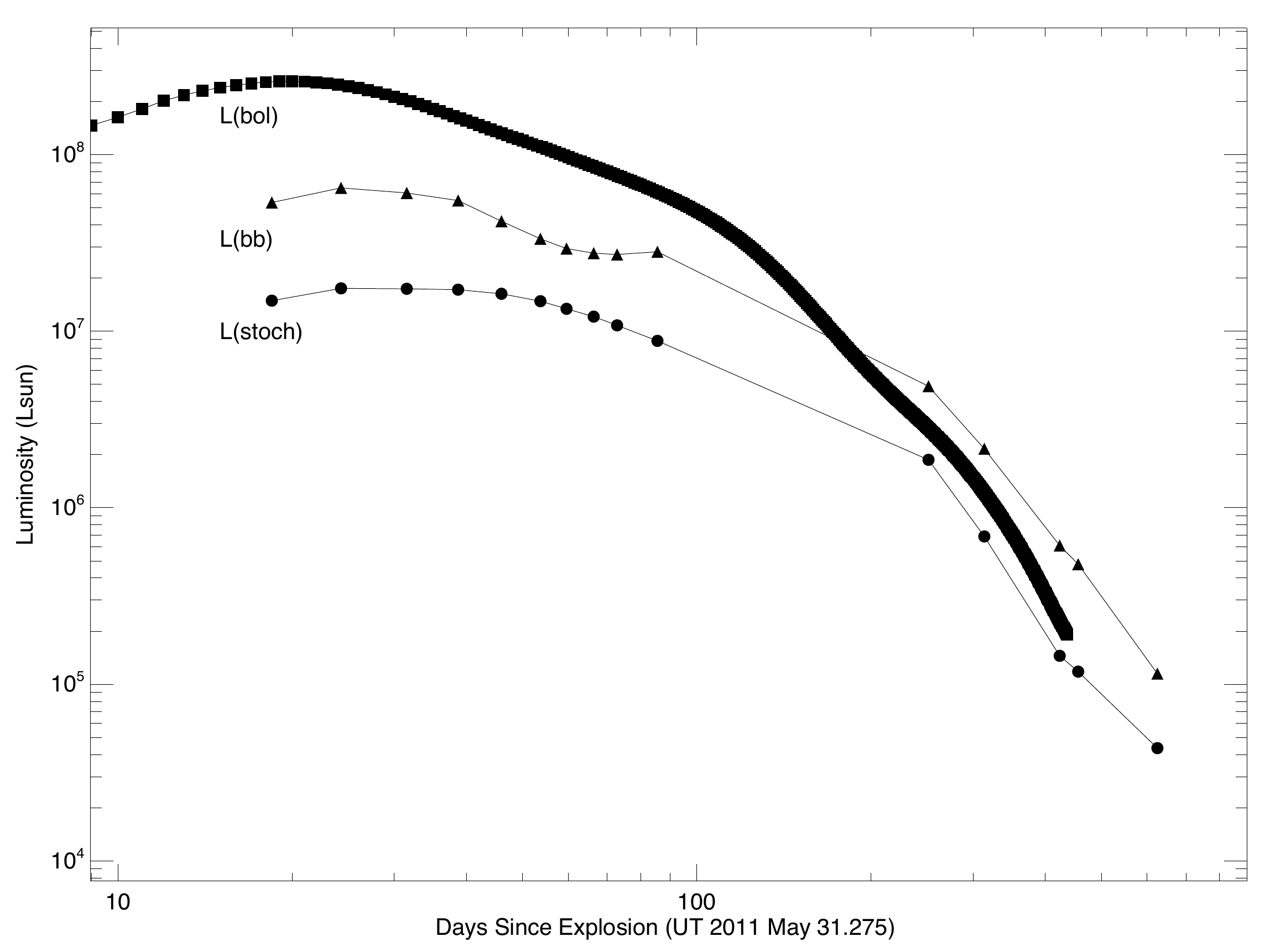}
\caption{Luminosity estimates for SN\,2011dh  as a function of time.
The stochastic heating case luminosity is computed as the sum $\nu*f_{\nu}(3.6)+\nu*f_{\nu}(4.5)$.
The thermal emission case luminosity is computed by fitting a blackbody curve to f$_{\nu}$(3.6) and f$_{\nu}$(4.5).
L(bol) sums over the UV-to-Vis part of the spectrum.
}
\label{fig:lumest}
\end{figure}

Assuming  dust in thermal equilibrium and ignoring the temperature spread,  the ratio $f_\nu(3.6)/f_\nu(4.5)$ yields a temperature.  Combined with the observed luminosity at 4.5$\mu$m, the latter leads to a total luminosity $L(bb)$, essentially an extrapolation of the two mid-IR bands to the full blackbody curve (Figure~\ref{fig:lumest}). Because 3.6$\mu$m carries additional contributions, $L(bb)$ underestimates the luminosity of a blackbody dominating at 4.5$\mu$m.

%%% \subsection{Stochastic Heating Case}

In the  case of stochastic heating, very small grains absorb single photons, radiate briefly at very high apparent temperatures and cool off rapidly \citep{dl01}.  Since the IR color cannot be associated with a temperature, and the spectrum is insufficiently constrained, we use $L(stoch)= \nu\,\times\, f_\nu(3.6)+\nu\,\times\,f_\nu(4.5)$ as a lower limit to $L(IR)$, even though 3.6$\mu$m is contaminated by the hot component at early times.

Both  $L(stoch)$ and  $L(bb)$ are  within a factor of 2 of being constant for the first 100\,days, and  differ by a factor ranging from 2 to 4 (Figure~\ref{fig:lumest}). They set a plausible range for the true $L(IR)$.   In the first 60 days, $L(bb)$\,  is a third of $L(bol)$\, which sums over the UV-to-Vis part of the spectrum.
%%%  the visible-light luminosity estimated from the  R band, and also plotted in Figure~\ref{fig:lumest}.  
$L(IR)/L(bol)$\, increases  substantially with time, so that even $L(stoch)/L(vis)$ approaches unity after 250 days. 
$L(IR)$\, is in the range of a few $10^7~L_\odot$ in the first month, and remains above $10^7~L_\odot$ for 100 days.  At peak, IR emission associated with the SN accounts for 2\% and 3\% of the M51 luminosity  at  3.6 and 4.5$\mu$m respectively 
\citep{dbe+05}.

The IR emission integrated  up to 90 days comes to $3.4\,\times\,10^{47}$\,ergs for $L(stoch)$ and to $10^{48}$\,ergs for $L(bb)$.  
%%%Correcting for the extrapolation from the visible and NIR as described in the next section reduces it by half.
The ratio of $L(bb)$ to $L(bol)$ summed over all epochs is $\approx{0.3\pm0.1}$, which translates to a total optical depth in the dust shell, $\tau_d$.
For this value of $\tau_d$ and the sizes derived above for inner and outer radii, total mass estimates for the shell, using either \cite{f+10} or \cite{d83}, would run well over 100 M$_{\odot}$, unrealistic for a circumstellar shell.  For the same $\tau_d$, cloud size would have to be an order of magnitude smaller to make the mass estimates consistent with stellar outflows, but the size would then clash with the timing arguments above. 
This points strongly to an interstellar cloud adding to the IR echo.
 
%%%  This increase in the  ratio cannot be due to variation in an optical depth term converting L(vis) to L(IR), but rather a result of the IR echo from both shock breakout flash and main fireball fading more slowly  than the  fireball itself. 

Under certain conditions, it may be possible for heating by the SBF to  dominate the early IR echo, namely until its IR emission surface has cleared $r_{sub}$ and the main light curve has risen significantly, a total of  $2\,t_{sub}+t_{rise}$.  If such conditions were to obtain, summing over the early IR echo would yield an approximate  lower limit to SBF UV-Vis energy, with the amount of underestimation depending on the SBF spectrum, and associated  losses to dust destruction, gas ionization, escape or other loss mechanisms.  
The examination of this possibility is deferred to a later paper, as it requires detailed modeling and analysis.

\subsection{Thermal Echo Modeling}

\begin{figure}[!ht] 
\centering
\includegraphics[width=\textwidth]{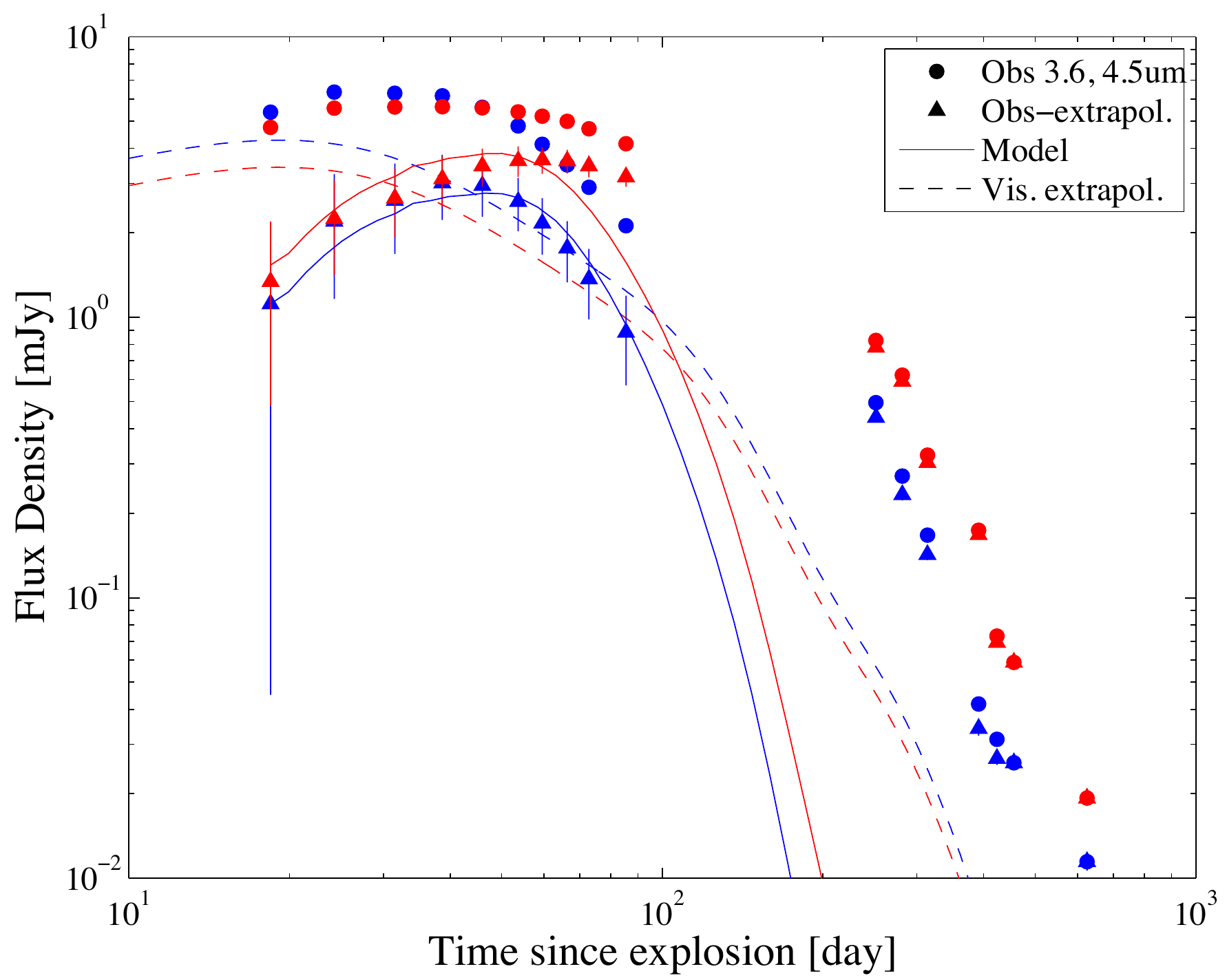}
\caption{Thermal echo model fit to SN\,2011dh (solid lines). Blue is 3.6$\mu$\,m and Red is 4.5$\mu$\,m. 
The Spitzer light curve data (circles), the visible light light curve extrapolated to
the infrared (dashed lines) and the difference (triangles) are also shown. At early time, the MIR emission 
can be described by the thermal echo. At late time, an additional dust heating mechanism is needed. 
% described with a model with parameters:Temperature=2000\,K, Emissivity(3.6) = 0.5, Emissivity(4.5) = 0.32 and L(peak) = 1e44 erg s$^{-1}$.
}
\label{fig:echo}
\end{figure}

The simple scaling arguments advanced above for IR echoes cannot account for all observables simultaneously. We therefore attempted to reproduce the Spitzer observations using simple IR echo models, with the following assumptions: the heating is done by a SBF  followed by the observed UV-Vis light curve; dust sublimates at 2000~K; the relative dust emissivity is $\epsilon(3.6\mu\,m)/\epsilon(4.5\mu\,m)=1.56$; L(peak) = $10^{44}$\,erg/s, comparable to the expected values in a Red Super Giant explosion (Nakar \& Sari 2010).   We extrapolated the visible spectrum to the IR assuming a Rayleigh-Jeans tail, and subtracted the extrapolation from the IR data (Figure~\ref{fig:echo}).
This highly uncertain extrapolation and the lack of coverage at $\lambda>4.5\mu$m affect all analysis of the data, including this modeling approach. 
Figure~\ref{fig:echo} shows the output of one model variant as an illustration rather than a preferred fit to the data.  In this variant, the dust has a flat radial profile around the SN, though the choice of radial gradient does not affect critically the predicted IR output. The  echo time dependence is most sensitive to $r_{sub}$, which in turn depends on L(peak).

Such simple models are  able to reproduce qualitatively the  shape of the IR light curves in the first 100\,days, though not the IR color evolution.   The models  turn over however before the observed plateau duration, and fall short of reproducing the late epoch data by more than an order of magnitude.  This difficulty  suggests additional emission components at later epochs besides the thermal echo.  
One possibility for an added component is emission from dust heated by direct contact with the shock wave which would still be propagating through the circumstellar envelope. The well-studied case of this phenomenon in SN\,1987A  \citep{dab+10} however shows increasing intensity between 6000 and 8000 days after explosion as opposed to the steep decline we observe here, possibly because of a different dust distribution.  
%%%  Another possibility  is direct emission from the photosphere of the SN itself, as suggested by the similar decay time of the $R$-band and mid-IR bands. One difficulty here  is the  essentially constant mid-IR color, though this may be due to strong emission lines in the mid-IR.  
Another possibility is CO fundamental band emission, invoked by  \cite{e+13}  as the explanation for 4.5$\mu$m excess even in the first 100 days.

Could this additional component simply arise in the pre-existing circumstellar dust heated by the shocked gas fireball?  The observed ratio of $L(IR)/L(bol)$ at late epochs would require an optical depth $>1$, and more importantly the observed T$\approx$500\,K and $L(bol)\approx\,10^6L_{\odot}$ would require the dust to be no further than 0.03\,pc from the heating source, smaller than $r_{sub}>0.1$pc.  This possibility is viable only in case of dust reforming after sublimation by SBF in the circumstellar material, which may take place on the time scale of hundreds of days \citep{g+11}, and would explain the increased optical depth after 200\,days.

Newly formed dust in the ejecta is another possibility that is widely discussed in the literature.  In a scenario such as depicted in Figure 9 of \cite{f+10}, new dust forming in the inner ejecta would be warmed by the shocked gas and obscure the rear half of that gas.  For SN2011dh, the shock speed of $2\times10^4$\,km\,s$^{-1}$ entails a shocked gas radius of 0.02\,pc, implying higher temperatures of interior dust than allowed by observations.  
For SN\,II, the review by Kosaza et al (2009)  models dust formation and IR emission increase over hundreds of years in core collapse SNe. Best known cases of ejecta-derived dust seem to occur in type IIn SN, and the dust temperatures run substantially higher there than in SN2011dh.  \cite{ms12} report on dust forming over a decade in SN 1998S (IIn)  whereas \cite{s+12} present evidence for dust forming over a three-month period  in SN 2010jl (IIn).  
%%% \cite{s+13} show evidence of circumstellar dust formation during pre-SN outbursts in SN\,2009ip (IIn).  
For illustration, the red continuum attributed to newly formed hot dust in SN 2006jc (Ibn)  reaches well below 1$\mu$m within tens of days, and is accompanied with  differential extinction of He II lines  \citep{s+08}.  The recent work by \cite{f+13} presents remarkable  results for a sample of IIn SNe, but their much higher and nearly constant luminosities over several years make them unlikely analogs for SN2011dh.
Ejecta-derived dust thus seems  an unlikely candidate for late excess emission in SN2011dh.  We plan to undertake detailed modeling in a later paper.

\section{Summary  and Discussion}

The IR emission from SN\,2011dh  evolves to lower color temperatures over the first three months after explosion, but also increases in importance compared to the visible light luminosity. It remains well above $10^7\,L_\odot$ during that period, and sums up to $\geq 3.4\times10^{47}\rm erg$, possibly an indication of the order of magnitude for the total UV--Vis energy in the shock breakout flash.  
%Depending on the duration of that flash,  its luminosity estimates range between a few $10^{8}L_\odot$\,  to over $10^{10}L_\odot$, which is comparable to the total luminosity of M51a. 
Higher cadence surveys should be sensitive to this bright flash (e.g. Zwicky Transient Facility, \citealt{k12}). 

Simple numerical models for the IR echo can reproduce qualitatively early mid-IR behavior, but then miss the 250--600\,day behavior, suggesting dust formation,  additional dust heating mechanisms, a nearby dense interstellar cloud, or significant line or band emission.

Thermal echoes could be effective at revealing SNe in highly obscured starbursts, where the reported rate of detection in the IR has been surprisingly low \citep{cmd+07}. Rather than search for the fast  rise of  extincted emission from young SNe, one should  look for thermal echoes which will be less extincted because of their spatial extent, and because part of the emitting surface will  approach the outer envelope of the starburst. 
%Z  The progenitor will have cleared out a region comparable to its Stromgren sphere before entering its late stages of evolution, stretching the thermal echoÕs peak delay to many days, and possibly months. 
Scaling from SN\,2011dh  at 8 Mpc, which remained brighter than 2 mJy at 3.6 and 4.5$\mu$m for 100 days, all SNe out to 20\,Mpc will have thermal echoes easily detectable with Spitzer even if they are 15 times less luminous. Most SN searches in starbursts were not looking for such slow profiles, and might have dismissed these light curves as dusty variable stars. Measuring SN  light echoes would constrain dust distribution models inside starbursts, and complement the picture we get from other tracers such as detailed infrared spectroscopy or radio molecular lines \citep{bbs+06, rhh+07, wpi+08}.

Acknowledgments:
We acknowledge the anonymous referee for helpful comments.  
This work was supported and based in part on observations made with the Spitzer Space Telescope, which is operated by the Jet Propulsion Laboratory, California Institute of Technology under a contract with NASA.
M.M.K. acknowledges generous support from the Hubble Fellowship and Carnegie-Princeton Fellowship. 
E.O.O. is incumbent of the Arye Dissentshik career development chair and is grateful to support by a grants from the Israeli Ministry of Science and the I-CORE Program.
Work by A.G.-Y. and his group is supported by grants from the BSF, ISF, GIF and Minerva, the EU/FP7 via an ERC grant, and the Kimmel award for innovative investigation.
%NEED PTF WORDING.

%\bibliographystyle{apj}
%\bibliography{ms}

%Chevalier & Fransson 2008; ApJ, 683, 185: cf08
%Rabinak & Waxman 2010; ArXiv e-prints arXiv:1002.3414: rw11
%Nakar & Sari 2010 ApJ, 725, 904: ns10
%Fox, O.D et al 2011, ApJ 741, 7: fcs+11
%Dwek, E. et al 2010, ApJ 722, 425: dab+10
%Weiler, K.~W, et al. 2007, ApJ, 671, 1959W: wwp+07 

\clearpage

\end{document}